# 探索資訊安全憑證專利技術

Abel C. H. Chen

Chunghwa Telecom Co., Ltd.

chchen.scholar@gmail.com; ORCID 0000-0003-3628-3033

## 摘要

由於如果資訊系統遭駭客入侵或攻擊，將可能造成個人資料外洩或重大的經濟損失。其中，資訊安全憑證是建構在公開金鑰基礎建設(Public Key Infrastructure, PKI)上的重要工具，可以應用在銀行交易、自然人憑證、區塊鏈、以及超文本傳輸安全協定(Hyper Text Transfer Protocol Secure, HTTPS)等服務。有鑑於此，本研究為分析資訊安全憑證創新技術的發展趨勢，採用專利資訊檢索系統，並且在系統中進行專利檢索和分析專利內容。本研究分別從單因子和雙因子的方式，觀察年度、技術領導企業、市場領導企業、以及主要應用領域之間的關聯，並且探討技術領導企業和市場領導企業在各個主要應用領域的專利佈局。

**關鍵字：資訊安全憑證、專利檢索、技術領導企業、市場領導企業**

# Exploring the Techniques of Information Security Certification

Abel C. H. Chen

Chunghwa Telecom Co., Ltd.

chchen.scholar@gmail.com; ORCID 0000-0003-3628-3033

## Abstract

If the information system is intruded or attacked by hackers, leaked personal data or serious economic loss may occur; the threats may be serious security problems. For security services, information security certification is built based on Public Key Infrastructure (PKI) to be an important tool for the services of bank transactions, natural person certificate, blockchain, and Hyper Text Transfer Protocol Secure (HTTPS). Therefore, this study uses Taiwan Patent Search System (TPSS) to find and analyze the contents of patents for obtaining the innovation reports of information security certification in Taiwan according to patents. This study considers the single-factor and two-factors to analyze the relationships of annuals, technology leaders, market leaders, and major applications for exploring the patent portfolios of technology leaders and market leaders in information security certification.

**Keywords: Information security certification, patent retrieval, technology leader, market leader**

# 1、 前言

近年來,駭客攻擊手法層出不窮,基礎設施遭駭客攻擊、企業遭勒索病毒集團挾持、個資外洩等資訊安全新聞報導時有所聞。例如,2022 年 10 月新聞「駭客入侵內政部?20萬筆宜蘭個資流出」報導有網友國外論壇 BreachForums 販售 2300 萬筆個資,並且已公開 20 萬筆真實資料(邱婉柔,2022;鄭閔聲,2022)。由此可知,資訊安全與個人及企業息息相關,並且有可能對人身財產和企業財產造成重大威脅。

在資訊安全系統中,加解密演算法主要可以分為對稱式加解密演算法(如:資料加密標準(Data Encryption Standard, DES)、進階加密標準(Advanced Encryption Standard, AES)等)和非對稱式加解密演算法(如:RSA (Rivest, Shamir, and Adleman)、橢圓曲線密碼學(Elliptic Curve Cryptography, ECC)等)。其中,公開金鑰基礎建設(Public Key Infrastructure, PKI)建設在非對稱式加解密演算法,可以產製私密金鑰、公開金鑰、以及簽發憑證等,並且應用在銀行交易、自然人憑證、區塊鏈、以及超文本傳輸安全協定(Hyper Text Transfer Protocol Secure, HTTPS)等服務(Dohare et al., 2022; Hewa et al., 2022),影響著我們每天的生活。

有鑑於此,研究資訊安全憑證創新技術的發展趨勢是個重要的研究議題。本研究將以專利資訊檢索系統為例,檢索資訊安全憑證相關專利,並且分析歷年發展趨勢、技術領導企業、市場領導企業、以及主要應用領域等。本研究的研究問題條列如下:

(1). 資訊安全憑證的熱點發展趨勢如何?

(2). 資訊安全憑證的技術領導企業有哪些?

(3). 資訊安全憑證的市場領導企業有哪些?

(4). 資訊安全憑證的主要應用領域有哪些?

本論文主要分為五個章節。第 2 節將介紹研究方法,說明專利檢索系統和專利檢索策略。第 3 節先從單因子的方式,各別從年度、第一申請人、第一專利權人、國際專利分類(International Patent Classification, IPC)號等面向,來觀察技術領導企業、市場領導企業、以及主要應用領域。第 4 節將採用雙因子的方式,各別從年度與第一申請人、年度與第一專利權人、第一申請人與國際專利分類號、以及第一專利權人與國際專利分類號等面向,探討技術領導企業和市場領導企業在各個主要應用領域的專利技術佈局。第 5 節將總結本研究的貢獻,並且討論未來研究發展方向。

2、 研究方法

為分析資訊安全憑證創新技術的發展趨勢，本研究採用專利資訊檢索系統(專利資訊檢索系統，2022)，並且在系統中進行專利檢索和分析專利內容，詳細說明如下。

在本研究中，以「憑證」作為關鍵字進行檢索，並且主要分析「發明專利」。由於本研究將分析專利權人的資訊，所以本研究主要分析「已核准發明專利」，已申請但尚未核准的發明專利不在本研究範圍內。有鑑於系統在每個月的 01 日、11 日、以及 21 日更新資料庫和系統，所以本研究的檢索日期是 2022 年 11 月 02 日，可以取得 2022 年 11 月 01 日（含）前核准的發明專利。檢索結果顯示從 1990 年到 2022 年期間，總共有 2,347 件憑證相關的發明專利被核准，並且本研究將在此檢索結果基礎上進行分析和討論。

3、 單因子分析

本節採用單因子分析的方式，從年度、第一申請人、第一專利權人、第一國際專利分類號各別因子來分析發展現況。其中，在專利申請時，同一件專利可能有一到多個申請人，通常第一申請人對該專利的技術掌握程度較高，所以本研究以第一申請人作為觀察技術領導企業的指標。此外，在專利被核准後，同一件專利可能有一到多個專利權人，假設專利鑑價後，通常第一專利權人對該專利的授權金可以擁有較高的比例，所以本研究以第一專利權人作為觀察市場領導企業的指標。最後，在專利審查的過程中，系統對同一件專利可能劃分一到多個國際專利分類號，通常第一國際專利分類號是最適合該專利的類別，所以本研究以第一國際專利分類號作為觀察主要應用領域的指標。

**(1) 熱點發展趨勢情況**

本研究採集 1990 年到 2022 年期間已核准的 2,347 件發明專利，依年度劃分結果如圖 1 和表 1 所示。根據本研究統計分析，2,347 件發明專利裡，申請到核准所需花費的時間，審核最久的專利花費 11 年，中位數和平均值分別為 3 年和 3.41 年。由此可知，仍有較多的專利需要 3 年以上的時間才會被核准。

由圖 1 呈現趨勢觀察，從 1998 年開始每年在資訊安全憑證申請發明專利數逐年上升，雖然在 2018 年有下降的情況，但因為 2018 年申請的專利到 2022 年 11 月時仍可能在審核中，所以申請發明專利數下降是仍有許多專利仍在審核中尚未核准所造成的。

此外，從核准發明專利數的趨勢觀察，可以發現每年核准發明專利數仍每年處於150件以上，是 2012 年之前每年核准發明專利數的數倍，反應出現在資訊安全憑證相對受到重視和被各個企業進行專利佈局。從表 1 熱點圖表示方式中，可以觀察 2013 年開始有比較多的發明專利被核准，並且在 2015 年和 2021 年核准發明專利數較多；其中，2015 年較多主要是陸續核准前幾年的專利申請案件，2021 年較多主要是 2020 年延遲的案件在 2021 年核准，整體趨勢仍維持在較多的核准發明專利數。

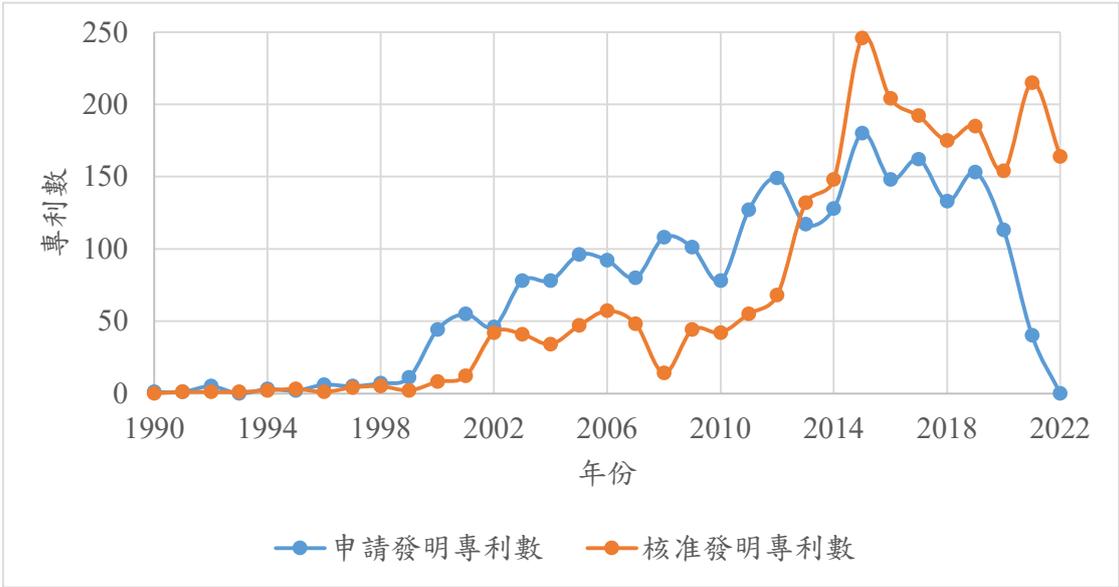

**圖 1 各個年度的申請發明專利數和核准發明專利數**

**(2) 技術領導企業情況**

有鑑於第一申請人對該專利的技術掌握程度較高，所以本研究以第一申請人作為觀察技術領導企業的指標。根據本研究統計分析，在檢索到的 2,347 件發明專利中，總共可以劃分到 783 個第一申請人，也就是平均每個第一申請人大約申請 3 件發明專利。然而，如果把核准發明專利數由高到低排序的話，會發現大部分專利集中在少數幾家企業，並且有 529 個第一申請人僅申請 1 件發明專利，呈現指數分佈。因此，觀察少數申請大量發明專利的企業更有助於了解資訊安全憑證產業的發展情況。

本研究主要取得核准發明專利數前十名的第一申請人，作為觀察技術領導企業的主要對象，如表 2 所示。其中，申請最多資訊安全憑證相關發明專利的技術領導企業是中華電信股份有限公司，之後依序是英特爾公司、高通公司、開曼群島商創新先進技術有限公司、微軟公司、萬國商業機器公司、阿里巴巴集團服務有限公司、財團法人

工業技術研究院、臺灣網路認證股份有限公司。在後續章節中將對這些技術領導企業進行深入討論，分析其專利交互授權和專利佈局情況。

表 1　各個年度的申請發明專利數和核准發明專利數

| 年份 | 申請發明專利數 | 核准發明專利數 |
| --- | --- | --- |
| 1990 | 1 | 0 |
| 1991 | 1 | 1 |
| 1992 | 5 | 1 |
| 1993 | 0 | 1 |
| 1994 | 3 | 2 |
| 1995 | 2 | 3 |
| 1996 | 6 | 1 |
| 1997 | 5 | 4 |
| 1998 | 7 | 5 |
| 1999 | 11 | 2 |
| 2000 | 44 | 8 |
| 2001 | 55 | 12 |
| 2002 | 46 | 42 |
| 2003 | 78 | 41 |
| 2004 | 78 | 34 |
| 2005 | 96 | 47 |
| 2006 | 92 | 57 |
| 2007 | 80 | 48 |
| 2008 | 108 | 14 |
| 2009 | 101 | 44 |
| 2010 | 78 | 42 |
| 2011 | 127 | 55 |
| 2012 | 149 | 68 |
| 2013 | 117 | 132 |
| 2014 | 128 | 148 |
| 2015 | 180 | 246 |
| 2016 | 148 | 204 |
| 2017 | 162 | 192 |
| 2018 | 133 | 175 |
| 2019 | 153 | 185 |
| 2020 | 113 | 154 |
| 2021 | 40 | 215 |
| 2022 | 0 | 164 |

表 2　核准發明專利數前十名的第一申請人

| 第一申請人 | 核准發明專利數 |
|---|---|
| 中華電信股份有限公司 | **121** |
| 英特爾公司 | **110** |
| 蘋果公司 | **102** |
| 高通公司 | **94** |
| 開曼群島商創新先進技術有限公司 | **75** |
| 微軟公司 | **70** |
| 萬國商業機器公司 | **56** |
| 阿里巴巴集團服務有限公司 | **43** |
| 財團法人工業技術研究院 | **37** |
| 臺灣網路認證股份有限公司 | **33** |

**(3)　市場領導企業情況**

　　有鑑於第一專利權人對該專利的授權金可以擁有較高的比例，所以本研究以第一專利權人作為觀察市場領導企業的指標。根據本研究統計分析，在檢索到的 2,347 件發明專利中，總共可以劃分到 784 個第一專利權人，並且有 526 個第一專利權人僅擁有 1 件發明專利，呈現指數分佈，大致與第一申請人的情況相似。由於專利核准後，如果專利沒有做轉讓或授權等相關變動的情況下，第一申請人和第一專利權人會是相同的。因此，第一申請人和第一專利權人的分佈情況相似是一般現象。然而，本研究發現在少數申請大量發明專利的企業（即本研究定義的「技術領導企業」）和少數擁有大量發明專利的企業（即本研究定義的「市場領導企業」）不同，值得深入分析與討論。

　　本研究主要取得核准發明專利數前十名的第一專利權人，作為觀察市場領導企業的主要對象，如表3所示。其中，擁有最多資訊安全憑證相關發明專利的市場領導企業是中華電信股份有限公司，之後依序是英特爾公司、蘋果公司、開曼群島商創新先進技術有限公司、高通公司、微軟技術授權有限責任公司、萬國商業機器公司、財團法人工業技術研究院、臺灣網路認證股份有限公司、惠普發展公司有限責任合夥企業。從核准發明專利數觀察，可以得知中華電信股份有限公司主要都是自主研發的技術，並且主要用來保護公司自身的產品，未轉讓和未授權給其他企業。而其他第一專利權人的企業相較於該企業擔任第一申請人時的核准發明專利數則略有增加的情況，也就是有通過購買或其他方式取得發明專利的轉讓或授權。

表 3 核准發明專利數前十名的第一專利權人

| 第一專利權人 | 核准發明專利數 |
|---|---|
| 中華電信股份有限公司 | **121** |
| 英特爾公司 | **112** |
| 蘋果公司 | **109** |
| <span style="color:red">開曼群島商創新先進技術有限公司</span> | <span style="color:red">**109**</span> |
| 高通公司 | **100** |
| <span style="color:red">微軟技術授權有限責任公司</span> | <span style="color:red">**88**</span> |
| 萬國商業機器公司 | **50** |
| 財團法人工業技術研究院 | **36** |
| 臺灣網路認證股份有限公司 | **33** |
| 惠普發展公司有限責任合夥企業 | **32** |

其中，核准發明專利數增加較多的企業是開曼群島商創新先進技術有限公司，並且其獲得核准發明專利的來源主要來自於阿里巴巴集團服務有限公司。深入分析企業間的關聯，開曼群島商創新先進技術有限公司是阿里巴巴集團服務有限公司的子公司之一，而阿里巴巴集團服務有限公司申請獲得核准的發明專利大部分都轉讓或授權給開曼群島商創新先進技術有限公司。因此，開曼群島商創新先進技術有限公司的核准發明專利數從 75 件增加到 109 件。

此外，微軟公司和微軟技術授權有限責任公司也有類似的作法。微軟技術授權有限責任公司是微軟公司的子公司之一，而微軟公司申請獲得核准的發明專利大部分都轉讓或授權給微軟技術授權有限責任公司。因此，微軟技術授權有限責任公司的核准發明專利數從 29 件增加到 88 件。

**(4) 主要應用領域情況**

有鑑於發明專利所屬的國際專利分類號是由審查機關設定，具有公信力，並且第一國際專利分類號是最適合該專利的類別，所以本研究以第一國際專利分類號作為觀察主要應用領域的指標。根據本研究統計分析，在檢索到的 2,347 件發明專利中，總共可以劃分到 89 個第一國際專利分類號，也就是平均每個第一國際專利分類大約申請 26 件發明專利。然而，如果把核准發明專利數由高到低排序的話，會發現大部分專利集中在少數幾個國際專利分類號，僅有 7 個第一國際專利分類號具有 26 件以上的發明專利，如表 4 所示。因此，觀察少數具有大量發明專利的國際專利分類號更有助於了解資訊安全憑證產業的主要應用領域情況。

表 4　核准發明專利數前十名的第一 IPC (IPC 國際專利分類查詢，2022)

| 第一 IPC | IPC 說明 | 核准發明專利數 |
| --- | --- | --- |
| G06F | 電子數位資料處理 | **622** |
| G06Q | 專門適用於行政、管理、商業、經營、監督或預測目的的數據處理系統或方法；其它類目不包含的專門適用於行政、管理、商業、經營、監督或預測目的的數據處理系統或方法 | **536** |
| H04L | 數位資訊之傳輸，例如電報通信 | **503** |
| H04W | 無線通訊網路 | **207** |
| G06K | 數據識別；數據表示；記錄載體：記錄載體之處理 | **74** |
| H04N | 影像通信，例如電視 | **46** |
| H01L | 半導體裝置；其他類目未包括的電固體裝置 | **28** |
| G11B | 基於記錄載體與轉換器之間之相對運動而實現的資訊儲存記憶體 | **22** |
| G07C | 時間登記器或出勤登記器；登記或指示機器之運行；產生隨機數；投票或彩票設備；其他類目不包括之核算裝置，系統或設備 | **21** |
| H04B | 傳輸 | **20** |

表 4 顯示具有最多資訊安全憑證相關發明專利的第一國際專利分類號是 G06F、G06Q、H04L、以及 H04W，對應主要應用領域分別是電子數位資料處理、數據處理系統或方法、數位資訊之傳輸、以及無線通訊網路。其他的第一國際專利分類號和主要應用領域依序是 G06K（數據識別）、H04N（影像通信）、H01L（半導體裝置）、G11B（資訊儲存記憶體）、G07C（時間登記器或出勤登記器）、H04B（傳輸）(IPC 國際專利分類查詢，2022)。

由資料顯示，資訊安全憑證主要是一種電子數位資料處理，所以有622件已核准發明專利劃分到 G06F。並且，資訊安全憑證多數應用在數據處理系統或方法，包含行政、管理、商業、經營、監督或預測等服務，所以有 **536** 件已核准發明專利劃分到 G06Q。此外，傳輸過程中也會需要使用到資訊安全憑證進行安全通訊，所以分別有 503 件、207件劃分到H04L、H04W。在雙因子分析章節中將對各個技術領導企業和各個市場領導企業在各個主要應用領域的專利佈局進行深入討論。

4、　雙因子分析

本節採用雙因子分析的方式，分別從技術領導企業的年度發展情況、技術領導企業的主要應用領域情況、以及市場領導企業的主要應用領域情況等三個面向進行討論。

**(1) 技術領導企業的年度發展情況**

為分析技術領導企業的年度發展情況，本研究從年度和第一申請人兩個因子來分析其關聯，並且以核准發明專利數前十名的第一申請人作為技術領導企業進行觀察和討論，如表 5 所示。在 1990 年到 2022 年期間已核准的 2,347 件發明專利，可以觀察到中華電信股份有限公司、英特爾公司、微軟公司、萬國商業機器公司、以及財團法人工業技術研究院都在 2005 年之前就開始在資訊安全憑證領域進行專利佈局，長期在資訊安全憑證的技術深耕。並且，由資料顯示中華電信股份有限公司在 2015 年之後加大投入在資訊安全憑證的技術研發，被核准發明專利數較早期提升數倍；此外，英特爾公司也在 2015 年到 2018 年期間加大投入在資訊安全憑證的技術研發，在這 4 年間增加 73 件已核准發明專利。然而，微軟公司、萬國商業機器公司則 2015 年後開始減少投入在資訊安全憑證的技術研發。此外，高通公司在 2013 年開始投入較多資訊安全憑證的技術研發，開始在資訊安全憑證領域做專利佈局。在 4.(2)節和 4.(3)節中將對這些技術領導企業在各個主要應用領域進行深入討論，分析各個企業在主要應用領域的發展情況。

值得關注的還有近年開始大量投入在資訊安全憑證技術研發的企業，包含開曼群島商創新先進技術有限公司和阿里巴巴集團服務有限公司。如 3.(3)節中描述阿里巴巴集團服務有限公司會將專利大部分轉讓或授權給開曼群島商創新先進技術有限公司，所以開曼群島商創新先進技術有限公司掌握了近幾年在資訊安全憑證領域的大部分發明專利。

**(2) 技術領導企業的主要應用領域情況**

為分析技術領導企業的主要應用領域情況，本研究從第一申請人和第一國際專利分類號兩個因子來分析其關聯，並且以核准發明專利數前十名的前十名第一申請人和第一國際專利分類號作為觀察目標進行深入討論，如表 6 所示。

以核准發明專利數最多的第一申請人中華電信股份有限公司為例，其在資訊安全憑證的專利佈局主要在 H04L（數位資訊之傳輸）、G06F（電子數位資料處理）、以及 G06Q（數據處理系統或方法）。因為，中華電信股份有限公司主要是電信業，並從事資通訊產品服務的開發。因此，中華電信股份有限公司在資訊安全憑證領域，主要著重在資料傳輸和處理，並且開發部分資料處理系統和平台。

表 5 核准發明專利數前十名的第一申請人在各年度的核准發明專利數

| 年份 | 中華電信股份有限公司 | 英特爾公司 | 蘋果公司 | 高通公司 | 開曼群島商創新先進技術有限公司 | 微軟公司 | 萬國商業機器公司 | 阿里巴巴集團服務有限公司 | 財團法人工業技術研究院 | 臺灣網路認證股份有限公司 |
|---|---|---|---|---|---|---|---|---|---|---|
| 1997 | 0 | 0 | 0 | 0 | 0 | 0 | 1 | 0 | 0 | 0 |
| 1998 | 0 | 0 | 0 | 0 | 0 | 0 | 0 | 0 | 0 | 0 |
| 1999 | 0 | 0 | 0 | 0 | 0 | 0 | 0 | 0 | 0 | 0 |
| 2000 | 1 | 0 | 0 | 0 | 0 | 0 | 0 | 0 | 1 | 0 |
| 2001 | 0 | 0 | 0 | 0 | 0 | 0 | 0 | 0 | 0 | 0 |
| 2002 | 3 | 1 | 0 | 0 | 0 | 0 | 0 | 0 | 0 | 0 |
| 2003 | 1 | 1 | 0 | 0 | 0 | 0 | 0 | 0 | 2 | 0 |
| 2004 | 1 | 1 | 0 | 0 | 0 | 0 | 4 | 0 | 2 | 0 |
| 2005 | 4 | 2 | 0 | 0 | 0 | 1 | 3 | 0 | 0 | 0 |
| 2006 | 4 | 5 | 0 | 0 | 0 | 4 | 5 | 0 | 2 | 0 |
| 2007 | 2 | 3 | 0 | 0 | 0 | 2 | 3 | 0 | 2 | 0 |
| 2008 | 2 | 1 | 0 | 0 | 0 | 0 | 0 | 0 | 1 | 0 |
| 2009 | 1 | 3 | 0 | 0 | 0 | 1 | 3 | 0 | 2 | 0 |
| 2010 | 1 | 5 | 0 | 2 | 0 | 5 | 5 | 0 | 1 | 0 |
| 2011 | 2 | 1 | 0 | 1 | 0 | 8 | 4 | 0 | 1 | 0 |
| 2012 | 2 | 0 | 0 | 3 | 0 | 11 | 6 | 0 | 0 | 0 |
| 2013 | 6 | 1 | 0 | 29 | 0 | 7 | 6 | 0 | 3 | 0 |
| 2014 | 5 | 6 | 10 | 7 | 0 | 16 | 3 | 0 | 4 | 0 |
| 2015 | 10 | 16 | 24 | 9 | 0 | 14 | 6 | 0 | 5 | 1 |
| 2016 | 8 | 16 | 19 | 9 | 0 | 1 | 2 | 0 | 3 | 3 |
| 2017 | 15 | 25 | 22 | 1 | 0 | 0 | 0 | 1 | 1 | 2 |
| 2018 | 9 | 16 | 15 | 6 | 0 | 0 | 0 | 0 | 3 | 4 |
| 2019 | 15 | 3 | 6 | 4 | 0 | 0 | 0 | 11 | 2 | 1 |
| 2020 | 5 | 2 | 4 | 7 | 9 | 0 | 0 | 25 | 0 | 8 |
| 2021 | 11 | 1 | 2 | 8 | 55 | 0 | 1 | 5 | 0 | 7 |
| 2022 | 13 | 1 | 0 | 8 | 11 | 0 | 2 | 1 | 1 | 7 |

此外，以核准發明專利數較多的第一申請人英特爾公司為例，其在資訊安全憑證的專利佈局主要在 G06F（電子數位資料處理）、H04L（數位資訊之傳輸）。因為，英特爾公司主要是半導體業，並從事處理器的開發。因此，英特爾公司在資訊安全憑證領域，更多是著重在資料處理，然後才是資料傳輸。高通公司也是半導體業，但更著重在無線通訊網路的技術，所以其在資訊安全憑證的專利佈局主要在 H04W（無線通訊網路）。另外，蘋果公司則是行動消費電子設備製造商，所以主要著重在無線通訊網路和設備的資料處理，所以其在資訊安全憑證的專利佈局主要在 H04W（無線通訊網路）、G06F（電子數位資料處理）。

表 6 核准發明專利數前十名第一申請人和第一 IPC 的專利數分佈情況

| 第一申請人 | G06F | G06Q | H04L | H04W | G06K | H04N | H01L | G11B | G07C | H04B |
|---|---|---|---|---|---|---|---|---|---|---|
| 中華電信股份有限公司 | 29 | 18 | 51 | 1 | 7 | 1 | | 2 | | |
| 英特爾公司 | 52 | 6 | 29 | 6 | 2 | 3 | 1 | | | 3 |
| 蘋果公司 | 29 | 10 | 14 | 31 | 1 | 7 | | | | 1 |
| 高通公司 | 6 | 1 | 21 | 63 | | | | | | 3 |
| 開曼群島商創新先進技術有限公司 | 24 | 35 | 8 | | 3 | | | 3 | | |
| 微軟公司 | 25 | 4 | 36 | 2 | 2 | | | | | |
| 萬國商業機器公司 | 23 | 1 | 28 | | | | | | | |
| 阿里巴巴集團服務有限公司 | 19 | 13 | 8 | 2 | 1 | | | | | |
| 財團法人工業技術研究院 | 6 | 7 | 12 | 5 | 3 | | | | 1 | |
| 臺灣網路認證股份有限公司 | 15 | 11 | 6 | | | | | 1 | | |

在開曼群島商創新先進技術有限公司和阿里巴巴集團服務有限公司在資訊安全憑證的專利佈局主要在 G06Q（數據處理系統或方法）、G06F（電子數位資料處理）。從資料統計分析觀察，可以發現這兩家公司在 G06Q（數據處理系統或方法）的專利佈局也較於其他企業來得多，其技術特徵主要落在資訊平台和資料處理系統的相關應用服務。

**(3) 市場領導企業的主要應用領域情況**

為分析市場領導企業的主要應用領域情況，本研究從第一專利權人和第一國際專利分類號兩個因子來分析其關聯，並且以核准發明專利數前十名的前十名第一專利權人和第一國際專利分類號作為觀察目標進行深入討論，如表 7 所示。其中，由於核准發明專利數最多的第一申請人和第一專利權人都是中華電信股份有限公司，並且中華電信股份有限公司沒有做專利轉讓或授權等權利變更申請，所以與 4.(2)節描述相同。在本節中將對第一申請人和第一專利權人變化較大的企業進行分析。

表 7  核准發明專利數前十名第一專利權人和第一 IPC 的專利數分佈情況

| 第一專利權人 | G06F | G06Q | H04L | H04W | G06K | H04N | H01L | G11B | G07C | H04B |
|---|---|---|---|---|---|---|---|---|---|---|
| 中華電信股份有限公司 | 29 | 18 | 51 | 1 | 7 | 1 | | 2 | | |
| 英特爾公司 | 52 | 6 | 29 | 7 | 2 | 3 | 1 | | | 3 |
| 蘋果公司 | 30 | 10 | 17 | 34 | 1 | 7 | | | | 1 |
| 開曼群島商創新先進技術有限公司 | 39 | 49 | 12 | | 4 | | | 3 | | |
| 高通公司 | 6 | 1 | 25 | 63 | | | | | | 3 |
| 微軟技術授權有限責任公司 | 43 | 7 | 33 | 1 | 2 | | | | | |
| 萬國商業機器公司 | 20 | 1 | 25 | | | | | | | |
| 財團法人工業技術研究院 | 6 | 7 | 12 | 5 | 3 | | | | 1 | |
| 臺灣網路認證股份有限公司 | 15 | 11 | 6 | | | | | 1 | | |
| 惠普發展公司有限責任合夥企業 | 25 | | 3 | 2 | | | | | | |

　　以核准發明專利數變化較大的第一專利權人開曼群島商創新先進技術有限公司為例，可以觀察到其在資訊安全憑證的專利佈局主要仍是在在 G06Q（數據處理系統或方法）、G06F（電子數位資料處理），但核准發明專利數顯著增加。其主要原因是阿里巴巴集團服務有限公司獲得核准發明專利數多數都轉讓或授權給開曼群島商創新先進技術有限公司，造成開曼群島商創新先進技術有限公司的核准發明專利數快速增加。此外，微軟公司和微軟技術授權有限責任公司之間也有類似的作法，由微軟公司申請和取得核准專利後，再把已核准發明專利轉讓或授權給微軟技術授權有限責任公司；因此，微軟技術授權有限責任公司以第一申請人和第一專利權人的核准發明專利數有顯著差異。

　　另外，英特爾公司、蘋果公司等企業在以第一專利權人的核准發明專利數也略有增加。此現象反應出國內外市場領導企業都積極在資訊安全憑證領域上做專利佈局。

## 5、 結論與建議

有鑑於資訊安全憑證的重要性，本研究通過專利資訊檢索系統採集 1990 年到 2022 年期間已核准的 2,347 件發明專利，並且對資訊安全憑證的熱點發展趨勢、技術領導企業、市場領導企業、以及主要應用領域進行深入討論與分析。由統計資料顯示，在資訊安全憑證領域的技術領導企業和市場領導企業是中華電信股份有限公司，其在資訊安全憑證領域，主要著重在資料傳輸和處理，並且開發部分資料處理系統和平台。另外，主要的技術領導企業還有英特爾公司、蘋果公司、高通公司、開曼群島商創新先進技術有限公司、微軟公司、萬國商業機器公司、阿里巴巴集團服務有限公司、財團法人工業技術研究院、臺灣網路認證股份有限公司等，資料顯示這些企業持續加大在資訊安全憑證領域的專利佈局。值得關注的還有開曼群島商創新先進技術有限公司和阿里巴巴集團服務有限公司，由於阿里巴巴集團服務有限公司將大部分的核准發明專利都轉讓或授權給開曼群島商創新先進技術有限公司，所以開曼群島商創新先進技術有限公司已成為近幾年資訊安全憑證發明專利最多的企業，值得後續深入追蹤分析。

## 參考文獻